\documentclass{elsart}

\usepackage{harvard}

\usepackage{graphicx}

\usepackage{amssymb}

\def\url#1{{\ttfamily\def\/{/\discretionary{}{}{}}#1}}

\begin{document}

\begin{frontmatter}
\title{Radio galaxy evolution:  what you can learn from a Brief Encounter}


\author[Blundell]{Katherine M.\ Blundell\thanksref{kmb}}, 
\author[Blundell]{Steve Rawlings}, 
\author[Blundell,Willott]{Chris J.\ Willott}, 
\author[Kassim]{Namir E.\ Kassim} \& 
\author[Perley]{Rick Perley}

\thanks[kmb]{E-mail: kmb@astro.ox.ac.uk}

\address[Blundell]{University of Oxford Astrophysics, Keble Road,
Oxford, OX1 3RH, UK}
\address[Willott]{Instituto de Astrofisica de Canarias, 38200 La Laguna, 
Tenerife, Spain}
\address[Kassim]{NRL, Code 7213, Washington DC, 20375-5351, USA}
\address[Perley]{NRAO, P.O.\ Box 0, Socorro NM, 87801-0387, USA }

\begin{abstract}

We describe the pitfalls encountered in deducing from classical
double radio source observables (luminosity, spectral index,
redshift and linear size) the essential nature of how these
objects evolve.  We discuss the key role played by hotspots in
governing the energy distribution of the lobes they feed, and
subsequent spectral evolution.  We present images obtained using
the new 74\,MHz receivers on the VLA and discuss constraints
which these enforce on models of the backflow and ages in
classical doubles.

\end{abstract}
\end{frontmatter}

\section{Introduction}
\label{intro}
\begin{figure}
\begin{center}
\includegraphics*[width=10cm]{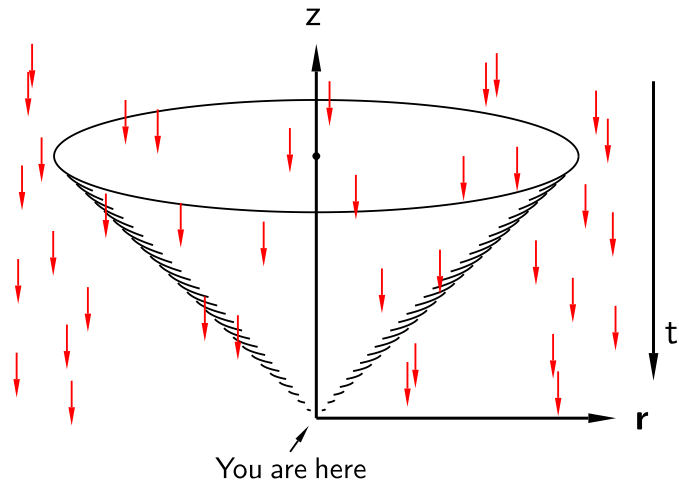}
\end{center}
\caption{A schematic illustration of our light-cone.  Each arrow
represents the timeline of a radio source, whose lifetime is
short compared to the Hubble time.  Only those arrows which
intercept our light-cone are those which we can observe.   The
point in a radio galaxy's lifetime when it is intercepted by our
light-cone is of course random. }
\label{fig:rain1}
\end{figure}

To identify the true nature of the ways in which radio galaxies
grow and evolve is not a trivial problem to tackle: we cannot
observe any radio galaxy at well-separated intervals throughout
its life.  We can only make observations of a particular radio
galaxy if it intercepts our light-cone.  Whether or not any radio
galaxy is intercepted by our light-cone, and at what point in its
life, is entirely random.  Fig.\ \ref{fig:rain1} schematically
illustrates the randomness of the sampling of any radio galaxy by
our light-cone.

Unavoidably imposed on the random sampling by our light-cone is
the effect of more selective sampling of the radio galaxy
population determined by the {\sl finding-frequency} of the
survey.  Different frequencies even within the radio regime, for
example 151\,MHz and 1.4 GHz, find different populations with
rather different selectivities: added complexities come from the
fact that across different radio-frequency regimes different
physical processes are responsible for the emission.  This is
schematically illustrated in Fig.\ 2.  At low frequency (e.g.\
151\,MHz) emission is largely isotropic while at GHz regimes
Doppler boosting preferentially selects objects with a small
angle to the light-of-sight.  In this way objects having low
intrinsic jet-powers may be found by the survey.  With GHz
surveys very young sources such as GPSs/CSOs may be found which
are synchrotron self-absorbed or free-free absorbed at lower
frequencies.

\begin{figure}[!t]
\begin{center}
\includegraphics*[width=10cm]{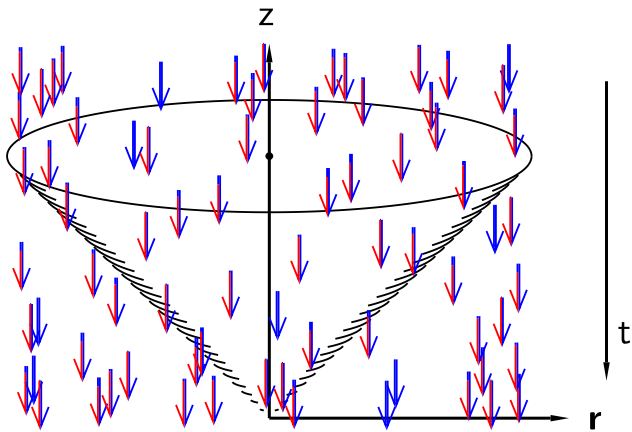}
\end{center}
\caption{A schematic illustration of our light-cone and its
interception by sources which might be seen according to the
survey finding-frequency.  The red or blue colours of the arrows
indicate the low or high frequencies at which the spectra of the
different objects peak.  Distinct, though not mutually exclusive,
samples will be obtained by different finding frequencies.  }
\label{fig:rain2}
\end{figure}

\begin{figure}
\begin{center}
\includegraphics*[width=10cm]{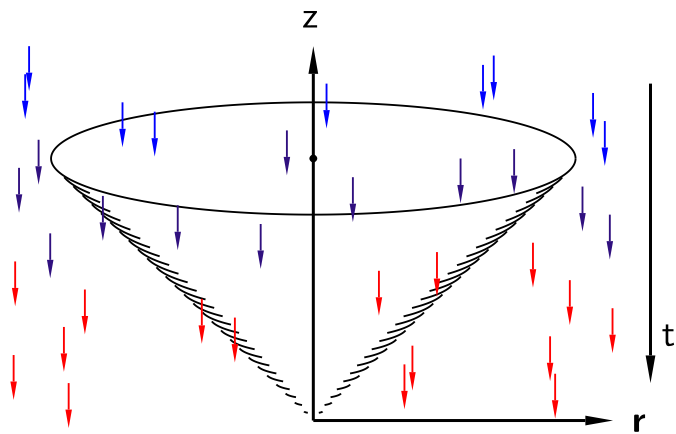}
\end{center}
\caption{A schematic illustration of our light-cone and the
inevitable $k$-correction which is convolved with the
finding-frequency of any survey: an object at redshift $z$ is
sampled on the basis of its rest-frame emission at $(1 + z)
\times$ the finding-frequency.  The red colours here indicate the
low-frequency at which the survey samples the emission.  The
increasingly blue colours of the arrows at higher-$z$ represent
the higher rest-frame frequencies at which they are sampled.  }
\label{fig:rain3}
\end{figure}

Unfortunately, the effect of a single finding-frequency varies as
a function of the redshift of the sources being surveyed.  For a
given survey finding-frequency $\nu_{\rm obs}$, the emission of a
given radio galaxy at redshift $z$ is sampled at rest-frame $(1 +
z) \times \nu_{\rm obs}$, as illustrated in Fig.\
\ref{fig:rain3}.  For objects characterised by steep spectra,
this makes their detection at high redshift harder for a given
observing frequency sensitivity.

\section{No direct evidence for luminosity evolution?}

{\em Direct} evidence for luminosity evolution in classical
doubles, for example the decreasing luminosity of a radio galaxy
with increasing size, could not be possible as Fig.\
\ref{fig:rain1} simply demonstrates: no radio galaxy has more
than a {\sl Brief Encounter} with our light cone.  We cannot make
well-spaced measurements along the time-axis of any given radio
galaxy.

Even if two properties such as luminosity and linear size might
be closely related within an individual source, it does {\em not}
follow that an anti-correlation will be manifested in the
luminosities and linear sizes of a sample of many radio sources.
The finding of Neeser et al.\ (1995) that the luminosities and
linear sizes of classical doubles from two complete samples, 3C
and 6C, are uncorrelated (which is substantiated by our own study
of three complete samples (3C, 6C and also 7C) in Blundell,
Rawlings \& Willott 1999, see \S\ref{sec:pd} therein) should {\em
not} be taken as the absence of evidence of a dependence of
source luminosity on source size/age.  Absence of evidence of
such a dependence would not in any case constitute evidence of
absence of such a dependence.  In order to establish a full
picture of the essential nature of how a radio galaxy evolves as
it gets older, it is essential to correctly understand the
influence of the sampling functions of the surveys and to
correctly measure the radio source observables (luminosity,
spectral index, redshift and linear size).  We now address these
issues.

\section{Sampling biases}

\begin{figure}
\begin{center}
\includegraphics*[width=13.5cm]{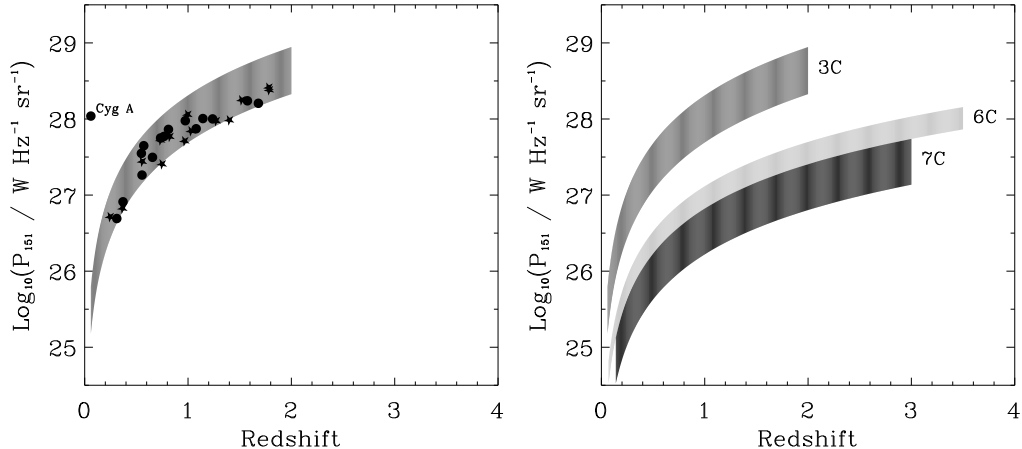}
\end{center}
\caption{{\em Left:} The coverage of the $P$--$z$ plane by the 3C
sample is shaded in grey; the filled circles represent sources
from the sample studied by Leahy, Muxlow \& Stephens (1987) and
the stars represent those sources studied by Liu, Pooley \& Riley
(1992) {\em Right:} Our study is based on three complete samples
selected at different flux-limits at 151\,MHz, a frequency
sufficiently low that Doppler biases are unimportant.  See
discussion on the finite widths of these regions in Blundell,
Rawlings \& Willott (1999). }
\label{fig:pz}
\end{figure}

The consequences of a flux-limit, together with the rarity of
classical doubles in the local Universe, mean that for a single
sample, such as 3C, there will be a tight correlation between the
luminosities ($P$) and redshifts ($z$) of the members of that
sample.  This is known as the $P$--$z$ degeneracy.  This is
illustrated in the left panel of Fig.\ \ref{fig:pz}, which shows
the distribution on the $P$--$z$ plane of sources from the
samples studied by Leahy, Muxlow \& Stephens (1989) and Liu,
Pooley \& Riley (1992).  Though these are two different samples,
they are both drawn from the 3C sample and subject to the $P$-$z$
degeneracy which afflicts the 3C or any other single flux-limited
sample.  Developing models of radio sources based on just such a
set of sources runs the risk of modelling the selection effect of
the $P$--$z$ degeneracy.

To circumvent such a problem, we have increased coverage of the
$P$--$z$ plane by using complete samples of radio sources which
are selected at lower flux-limits (Rawlings et al.\ 1998).  While
the flux-limit of the 3C sample is 12\,Jy at 151\,MHz, the
flux-limit of the faintest sample (7C) is 0.5\,Jy.  The
substantial improvement in coverage of the $P$--$z$ plane
achieved in this way is illustrated by the right panel in Fig.\
\ref{fig:pz}.

\section{Measurement of source properties}

\subsection{Spectral index}
\label{sec:spix}
\begin{figure}
\begin{center}
\includegraphics*[width=10cm]{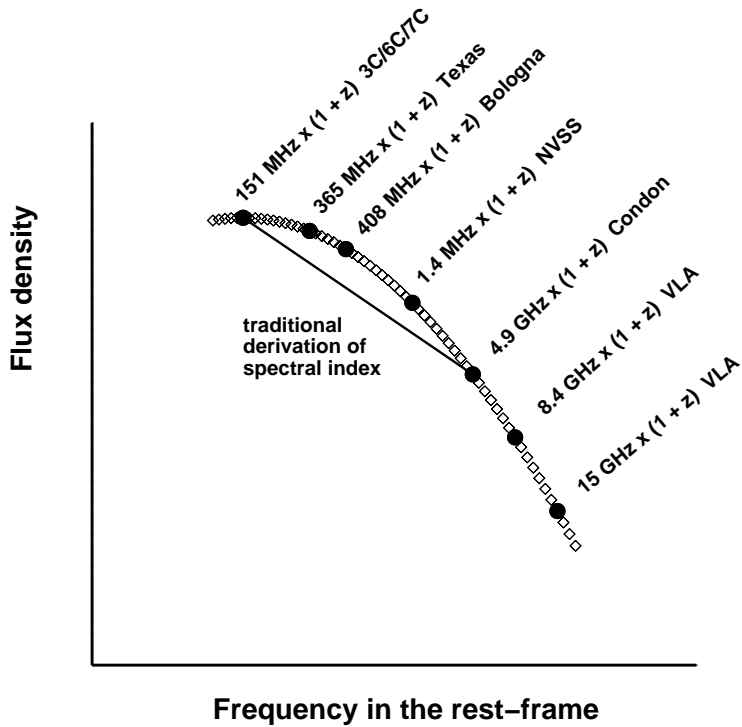}
\end{center}
\caption{Figure to illustrate the integrated spectral indices of
radio sources.  The traditional measurement of spectral index
(the gradient of the curve on a $\log$ flux-density {\em v}.\
$\log {\rm frequency}$ plot) assumes that the spectrum follows a
simple power-law.  A correct measurement of spectral index
involves fitting for the curvature of the spectrum and correcting
for the redshifting of the emission. }
\label{fig:spix}
\end{figure}

The integrated spectrum of a classical double radio source is
usually either concave (see Fig.\ \ref{fig:spix}) or described by
a simple power-law.  In order to properly measure the spectral
index at a common rest-frame frequency, it is necessary to fit
the curvature of the spectrum and perform a $k$-correction to
take account of the redshifted sampling of the spectrum.  We
performed spectral fitting and $k$-corrections as described in
Blundell, Rawlings \& Willott (1999) to all members of the 3C, 6C
and 7C complete samples used in our study, after obtaining from
the literature flux-density measurements for each object at 365
MHz, 408 MHz, 1.4 GHz and 4.86 GHz in addition to measurements
from our own VLA observations and from the survey measurements at
151\,MHz.  Hereafter in this paper we refer to the instantaneous
slope of the $\log$ flux-density -- $\log$ frequency plot
evaluated at rest-frame 151\,MHz as the spectral index at
151\,MHz or $\alpha_{\rm lo}$, and evaluated at rest-frame 5\,GHz
as the spectral index at 5\,GHz or $\alpha_{\rm hi}$.

\subsection{Luminosity}
Luminosity is not as fundamental a property of a radio source as
jet-power $Q$ is.  This is because in addition to depending on
the details of the environment into which the source is
expanding, luminosity also depends on the frequency regime in
which the luminosity is being evaluated and on the age of the
source, even if the bulk kinetic jet-power remains constant.  In
the context of comparisons between radio-loud and radio-quiet
quasars we discuss how luminosity at GHz frequencies is a poor
indicator of the underlying jet-power (Blundell \& Rawlings, in
prep.); this is especially but not exclusively true for those
objects having significantly Doppler boosted core emission.  For
the study we describe here, all luminosities are evaluated at
rest-frame 151\,MHz, following the spectral fitting procedure
mentioned in \S\ref{sec:spix}.

\subsection{Redshift}
Many previous studies of samples of radio sources have required
the use of redshift estimators for a significant fraction of the
sample (e.g. Dunlop \& Peacock 1990). The most common estimator
employs the $K$--band (2.2 $\mu$m) magnitude.  Use of this
estimator is motivated by the small spread in the near-infrared
Hubble Diagram ($K$--$z$ relation) for radio galaxies in the 3C
sample (Lilly \& Longair 1984; Best, Longair and R\"{o}ttgering
1998). It is now known that simple extrapolation of the 3C
relation to higher redshifts and/or to fainter radio samples
(which contain fewer radio luminous sources at any given
redshift) leads to both large uncertainties in the estimated
redshifts, and to systematic biases.  For the sources in the 3C,
6C and 7C samples used in our study, the redshifts are almost
exclusively spectroscopic.  In the very few cases where we have
been unable to secure spectroscopic redshifts we have estimated
redshifts by obtaining multi-colour (optical and near-infrared)
information and fitting spectral energy distributions (SEDs) with
template galaxy spectra (e.g. Dunlop \& Peacock 1993; Willott et
al.\ {\em in prep}).

\subsection{Linear size}
Angular sizes of radio sources were measured by taking the sum of
the separation from the core to each hotspot or, in the absence
of a core, the hotspot-to-hotspot separation.  Angular sizes were
converted into linear sizes in the usual way (e.g.\ Carroll,
Press \& Turner 1992) for three different cosmologies.  

\section{Salient correlations and non-correlations}
\subsection{The $z$-- $\alpha_{\rm lo}$ correlation}
\begin{figure}
\begin{center}
\includegraphics*[width=10cm]{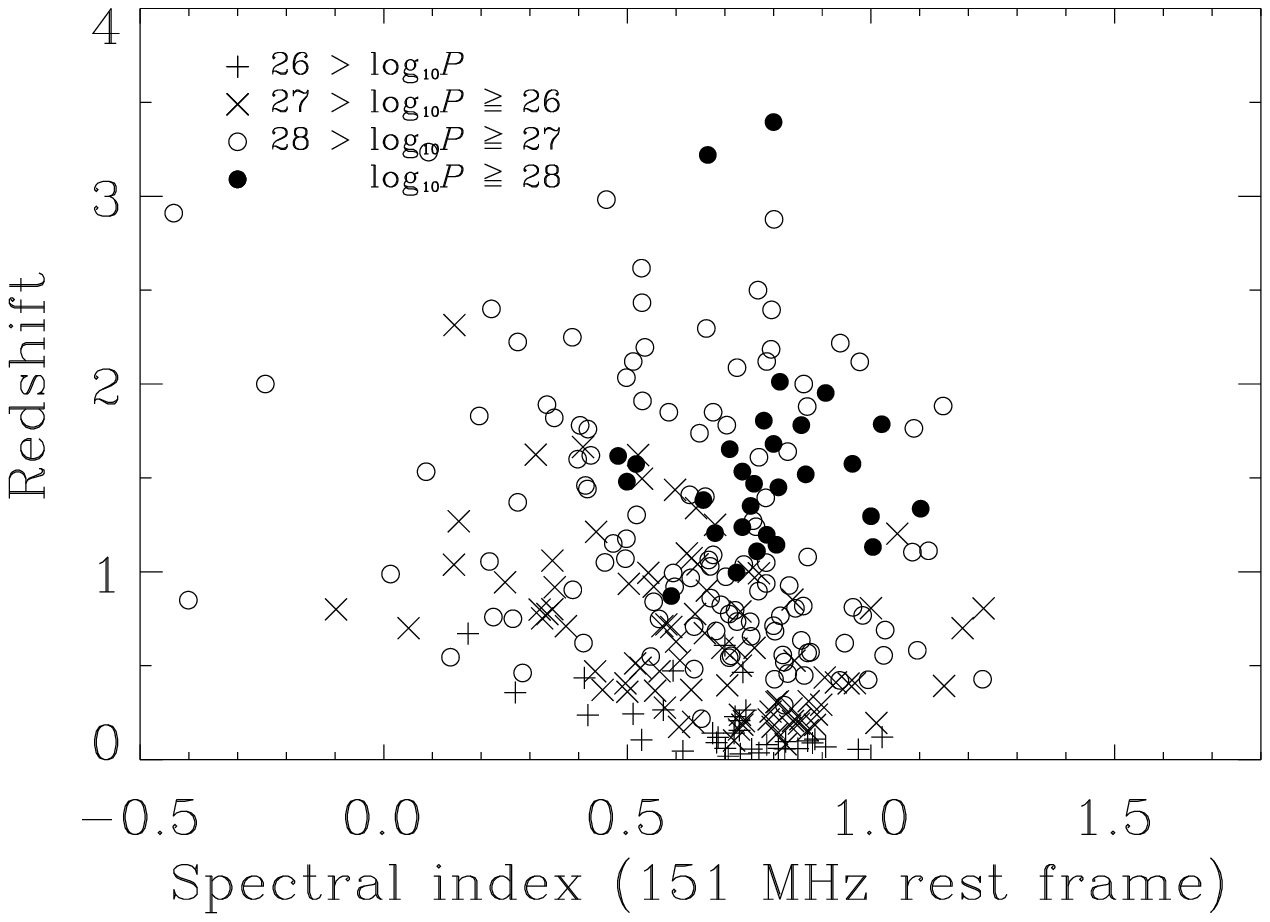}
\includegraphics*[width=10cm]{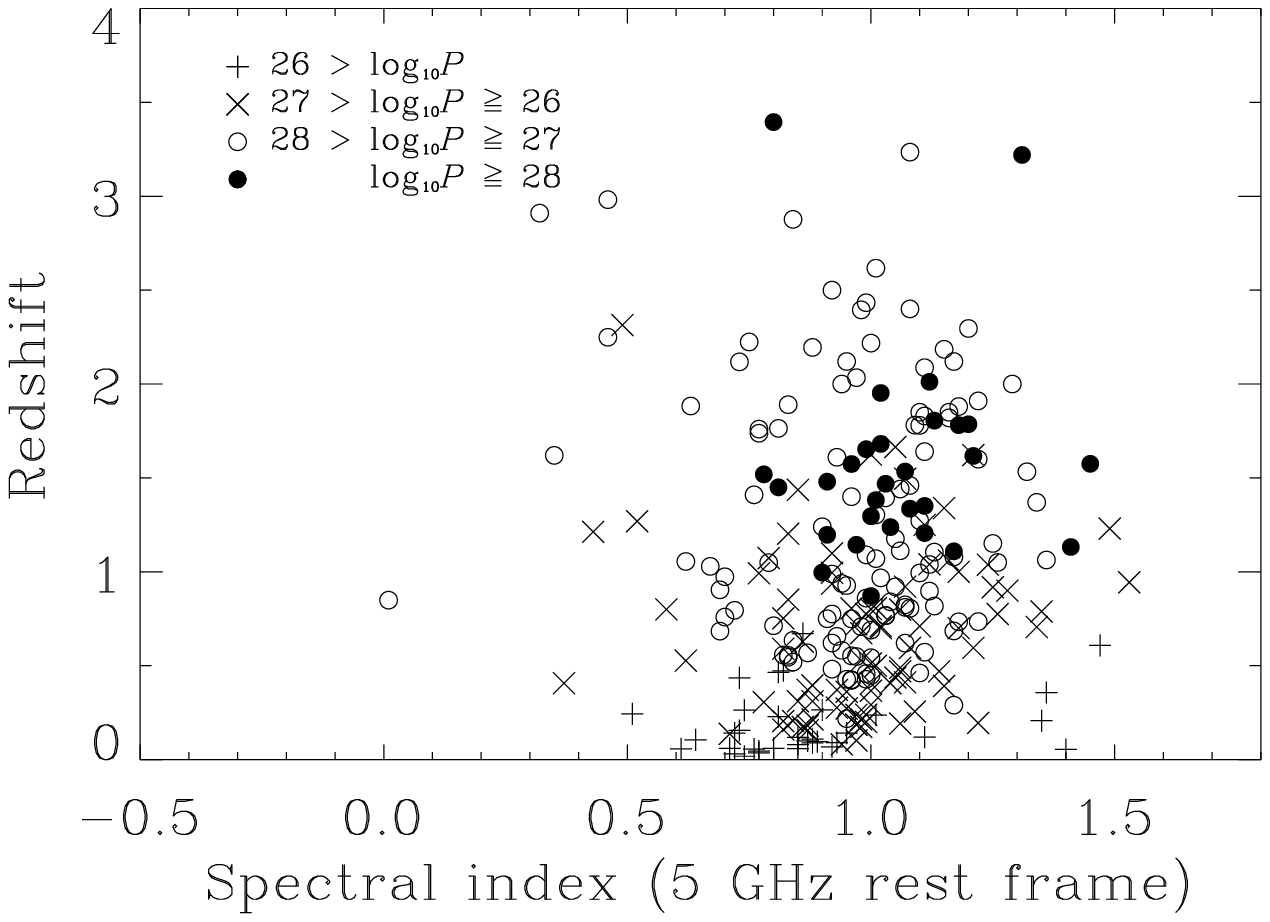}
\end{center}
\caption{Upper panel: redshift {\em v} spectral index
evaluated at 151\,MHz in the rest-frame.  Lower panel: redshift
{\em v} spectral index evaluated at 5\,GHz in the rest-frame.}
\label{fig:za}
\end{figure}

The low-frequency spectral index $\alpha_{\rm lo}$ (evaluated
at 151\,MHz in the rest-frame) is less subject to the effects of
synchrotron and inverse Compton losses off the CMB than
higher-frequency spectral indices ($\alpha_{\rm hi}$) evaluated
in the GHz regime.  As such, $\alpha_{\rm lo}$ is a closer
indicator of the energy distribution as initially injected than
is $\alpha_{\rm hi}$.  We find that there is no dependence on
redshift of $\alpha_{\rm lo}$, as shown in Fig.\ \ref{fig:za}
(upper panel).

\subsection{The $z$--$\alpha_{\rm hi}$ correlation}
In contrast with the picture in \S5.1 an increased dependence of
GHz spectral index is seen with redshift in Fig.\ \ref{fig:za}
(lower panel).  This points to the increasing importance of
inverse Compton losses off the CMB with increasing redshift,
which was first pointed out by Krolik \& Chen (1991).

\subsection{The $D$--$\alpha_{\rm lo}$ correlation}
Fig.\ 7 shows that the greater the linear size of a radio source
the steeper is its low frequency spectral index.  This was the
strongest correlation we found out of all of those we examined
for the set of complete samples under study.  We interpret this
as being a consequence of adiabatic expansion losses and telling
us that the magnetic fields decrease as the lobes expand, for the
following reason.  (We consider and reject other possibilities in
Blundell, Rawlings \& Willott 1999.)  For a fixed observing
frequency $\nu$, the relationship between the Lorentz factor
($\gamma$) of the particles giving rise to that emission and the
ambient magnetic field is given by:
\begin{equation}
\label{eq:lorentz}
\gamma = \biggl(\displaystyle\frac{m_{\rm e}}{eB} 2\pi \nu \biggr)^{
          \frac{1}{2}},
\end{equation}
where $e$ is the charge on an electron and $m_{\rm e}$ is its
rest-mass.  Thus as the magnetic field decreases, higher-$\gamma$
particles will be responsible for radiation at that fixed
observing frequency.  Since for lobe emission the spectrum of the
energy distribution of particles is steep and frequently curved
these higher-$\gamma$ particles will be from a steeper part of
the energy distribution, and hence a higher rest-frame 151\,MHz
spectral index will be observed.

\begin{figure}
\begin{center}
\includegraphics*[width=10cm]{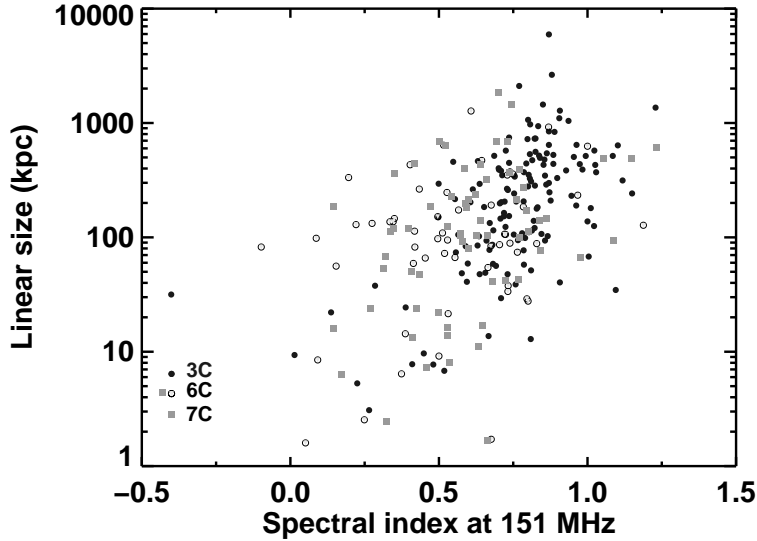}
\end{center}
\caption{The linear sizes of the radio sources plotted against
their spectral indices evaluated at 151\,MHz in the rest-frame.
}
\label{fig:da}
\end{figure}

\subsection{The $P$--$\alpha_{\rm lo}$ correlation}
\label{sec:pa}
A correlation between the luminosities of classical doubles and
their spectral indices was first suggested in 1960 by Heeschen,
and more recently by Laing \& Peacock (1980).  We find this
correlation to be present for the combination of 3C, 6C and 7C
complete samples in the sense that the more powerful the radio
source, the steeper is its low-frequency spectral index.

\begin{figure}
\begin{center}
\includegraphics*[width=10cm]{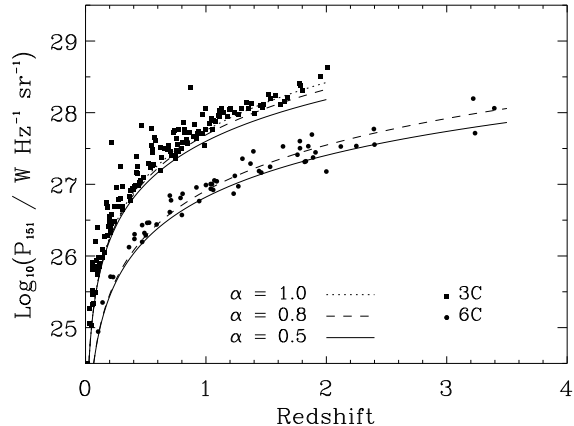}
\end{center}
\caption{The luminosities of the radio sources plotted against
their redshifts, with the flux-limits for the samples overlaid
for different assumed values of the low-frequency spectral index
for the 3C and 6C samples; the 7C data are omitted for clarity
but may be seen in figure 7 of Blundell, Rawlings \& Willott
(1999).  It can be seen that those 3C sources with $P >
27.5\,{\rm W\,Hz^{-1}\,sr^{-1}}$ all avoid the region between the
$\alpha = 0.5$ flux-limit and the $\alpha = 0.8$ flux-limit.
There is no such avoidance seen for the lower-power 6C sources
over the same redshift range. }
\label{fig:pa}
\end{figure}

One way of depicting the $P$--$\alpha$ correlation is to show for
the $P$--$z$ plane the effects of low-frequency spectral index on
the flux-limits and how sources from the different complete
samples cluster around the flux-limits (see Fig.\ \ref{fig:pa}).

We have modelled this $P$--$\alpha$ correlation as coming from a
dependence of the steepness of the energy distribution of
particles injected into the lobe on jet-power.  We do not presume
that the injection spectrum is universally given by the canonical
value of $\alpha = 0.5$.  We model the steepening of the energy
distribution of particles as being controlled largely by
classical synchrotron loss suffered in the high magnetic field of
the hotspot.  We specifically invoke that the hotspot magnetic
fields are higher in those sources with higher jet-powers (as
detailed in Blundell, Rawlings \& Willott 1999).  This point is
discussed further by De~Young (these proceedings).  We choose to
normalise this dependence of hotspot magnetic field on jet-power
by assuming equipartition (of the energy in the particles and the
energy in the magnetic field) in the hotspot (for which there is
good evidence in the case of Cygnus A, Harris et al.\ 1994).
We discuss further the role of the hotspot in \S\ref{sec:model}.

\subsection{The $D$--$z$ correlation}
\label{sec:Dz}
The `linear-size evolution' of radio sources has been known for
many years (Kapahi et al.\ 1987, Barthel \& Miley 1988).  We find
for our combination of complete samples selected at low frequency
that the decrease of linear size with increasing redshift is
quite mild, in the sense that the parameterisation of linear size
at constant luminosity and spectral index is given by $D \propto
(1 + z)^{-0.96}$.  We return to the physical mechanism
responsible for linear-size evolution of classical double radio
sources in \S\ref{sec:model}.

\subsection{The $P$--$D$ plane}
\label{sec:pd}

The $P$-$D$ plane shows that there is no strong correlation
between the luminosities and linear sizes of radio sources in the
combination of complete samples.  Given the sampling functions
alluded to at the start of this paper which influence the nature
of the objects which make it into our surveys, it should {\em
not} be claimed that the absence of a correlation between $P$ and
$D$ in any way might imply that for any given source as $D$
grows, $P$ will not evolve.

\begin{figure}
\begin{center}
\includegraphics*[width=10cm]{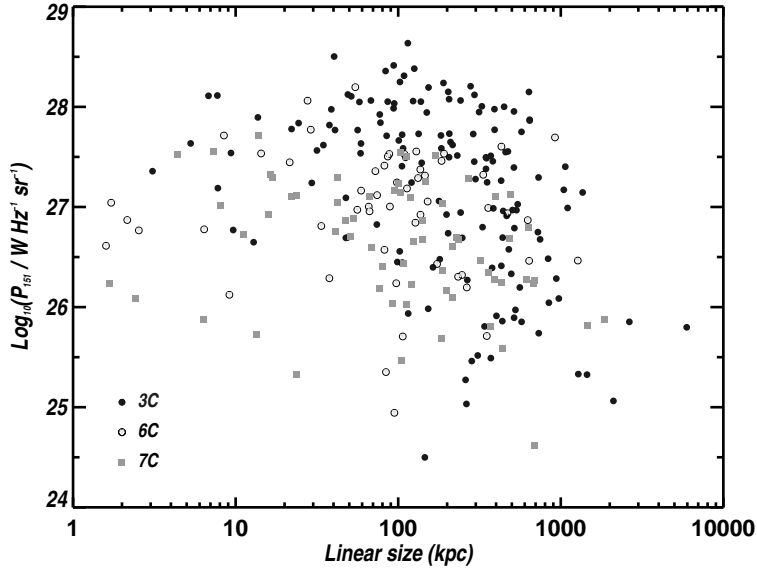}
\end{center}
\caption{The linear sizes of the radio sources from our complete
samples versus their luminosities.  No strong correlation is seen
between these properties for {\sl samples} of radio galaxies.  }
\label{fig:pd}
\end{figure}

\section{Refining models for luminosity evolution}
\label{sec:model}

We have formulated a model for evolving radio galaxies which
reproduces the above dependencies, and others, and this is
detailed in Blundell, Rawlings \& Willott (1999).  This model is
the first model for radio galaxy evolution which incorporates the
role played by the hotspot.  This role has two aspects. 

 {\em First}, the magnetic field of the hotspot causes
significant radiative losses on synchrotron particles during
their dwell-time in the hotspot, in the period of time before
they are injected into the lobe.  We were led to this as
discussed in \S\ref{sec:pa} by our observation that the radio
sources in our complete samples seemed to strongly indicate that
the more luminous a radio-source was, the steeper its {\sl
low-frequency spectrum} was (for example, when evaluated at
rest-frame 151\,MHz).  This rest-frame frequency regime is less
affected by synchrotron and inverse Compton losses than the GHz
regime and thus more closely informs us on the energy
distribution of particles {\sl as initially injected} into the
lobe of a radio-source.  Modelling a steeper spectrum in a more
powerful source came naturally out of invoking stronger magnetic
fields (hence stronger radiative losses) in sources with higher
jet-powers.  By equating the jet-thrust and the pressure in the
compact hotspot, and invoking equipartition in the hotspot, we
find that the magnetic energy density in the hotspot is
proportional to the bulk kinetic power transported in the jet.

{\em Second}, as a plasma element expands out of a compact
hotspot into a lobe whose pressure is considerably lower (and
continues to become lower as the radio lobe expands and gets
older) it suffers enhanced adiabatic expansion losses compared to
the expansion losses suffered by that element of plasma once it
continues to dwell in the lobe.  This is consistent with
observations of classical doubles which invariably show highly
compact hotspots --- embedded towards the outermost edges of the
smooth, low surface-brightness, extended emission which comprises
the lobe --- albeit with a bewildering menagerie of shapes and
structures (Leahy et al.\ 1997, Hardcastle et al.\ 1997  \& Black
et al.\ 1992).

\begin{figure}
\begin{center}
\includegraphics*[width=14cm]{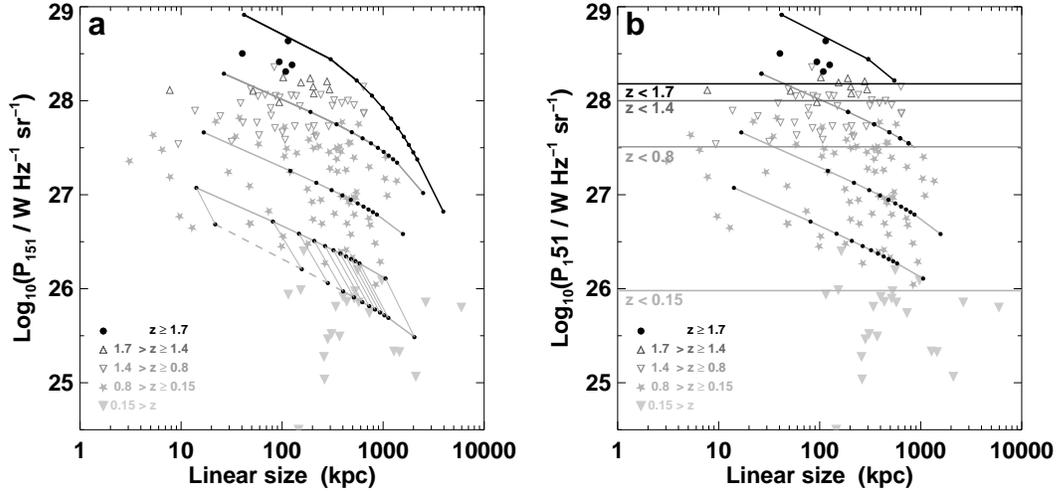}
\end{center}
\caption{Overlaid on the `$P$--$D$' plane for the 3C sample in
{\em \bf a} are model tracks tracing out the evolution in
luminosity and linear size of four example radio sources, with
from top to bottom $Q = 5 \times 10^{39}$ W at $z = 2$, $Q = 1
\times 10^{39}$ W at $z = 0.8$, $Q = 2 \times 10^{38}$ W at $z =
0.5$ and $Q = 5 \times 10^{37}$ W at $z = 0.15$.  Small dots lie
on all tracks to indicate the (rest-frame) ages 0.1, 1, 10,
20\ldots 100, 200 Myr.  The dashed line indicates how the lower
track luminosity reduces by $<$ half an order of magnitude if the
ambient density becomes an order of magnitude lower.  In {\em \bf
b} the horizontal lines represent the luminosities at which the
3C flux-limit of 12 Jy takes its effect at the different
redshifts indicated.  The same tracks as in {\em \bf a} are
plotted but here in {\em \bf b} they are truncated when they
reach the flux-limit.  Source expansion in these plots is
described by the model of Falle (1991).  In his model higher
jet-power sources expand more quickly; thus these objects fall
through the flux-limit at a younger age.  The luminosity
evolution is as described by the model of Blundell, Rawlings \&
Willott 1999, which incorporates the role of the hotspots. }
\label{fig:tracks}
\end{figure}

The earliest (Scheuer 1974; Baldwin 1982) to the most recent
(Kaiser, Dennett-Thorpe \& Alexander 1997; Blundell, Rawlings \&
Willott 1999) models of radio source evolution, with reasonable
assumptions about the environments into which radio sources
expand, predict that the luminosity of any individual radio
source decreases as time --- or age --- increases.  This
luminosity decline is enhanced when the role of the hotspot is
included and illustrative `tracks' across the $P$--$D$ plane of
radio sources with given jet-powers and redshifts are shown in
Fig.\ \ref{fig:tracks}.

A combination of the dramatically declining luminosity-with-age
of the high-$Q$ sources, their scarcity in the local Universe,
together with the harsh reality of survey flux-limits means that
very powerful sources with large linear sizes are rarely seen.
This resolves the old problem first posed by Baldwin in 1982 that
there is a dearth of large {\sl and} powerful classical doubles.
This problem had been a puzzle: intuitively [and as later
modelled by Falle (1991)] the more powerful radio-sources were
expected to expand faster than those with lower jet-powers, so
the laws of probability alone suggest it should be easy to find
large, powerful sources.  But they are not: sources with $P >
10^{27}\,{\rm W\,Hz^{-1}\,sr^{-1}}$ and with $D > 1\,{\rm Mpc}$
are extremely rare in existing surveys (see Fig.\ \ref{fig:pd}).

\section{The youth-redshift degeneracy}

In a recent letter to {\em Nature} (Blundell \& Rawlings 1999a)
we demonstrated how declining luminosity tracks in combination
with the consequences of applying a flux-limit at different
redshifts would inevitably give rise to an effect which we termed
the `youth-redshift degeneracy': the fact that in a given
flux-limited sample, those sources at higher redshift must be
younger than their low redshift counterparts.  This effect
applies very strongly to powerful sources when the role of the
hotspot is incorporated into the model $P$--$D$ tracks, as
described in \S6.

The youth-redshift (YZ) degeneracy means it is necessary to be
careful about the correct interpretation of `trends with
redshift'.  In many cases it is wrong to attribute such trends to
changes in {\em cosmic epoch} when the effects arise because of
{\em source age}.  We consider briefly five such examples of
effects which occur because the high redshift sources are young
sources, and hence are being observed not long after the
jet-triggering event (which is possibly a galaxy-galaxy merger,
Sanders et al 1988).

{\bf Linear-size evolution and the YZ-degeneracy} The linear size
evolution observed in low-frequency flux-limited samples of
classical double radio-sources (Kapahi et al.\ 1987, Barthel \&
Miley 1988, Blundell, Rawlings \& Willott 1999) arises because
the high--$z$ sources are younger, hence tend to be shorter.  It
is the positive dependence on jet-power of the rate at which the
lobe-lengths grow (Falle 1991) which contributes to the linear
size evolution being as mild as it is (see \S\ref{sec:Dz} and
Blundell, Rawlings \& Willott 1999).

{\bf Structural distortion and the YZ-degeneracy} Barthel and
Miley (1988) had suggested that higher redshift environments are
denser and more inhomogeneous than at low redshift since they
found increased distortion in the structures of their high-$z$
sample of steep-spectrum quasars compared with their low-$z$
sample.  Sources which are younger may have the passage of their
jets considerably more disrupted where there is a higher density
and greater inhomogeneity in the ambient post-recent-merger
environment.  A general trend of denser inter-galactic
environments at high-$z$ cannot be inferred from their result.

{\bf The alignment effect and the YZ-degeneracy} Where the
alignment effect of optical light along radio-jet axes is caused
by star-formation, it will be more easily triggered close in to
the host galaxy or within the product of a recent merger
[assuming this is the jet-triggering event (Sanders et al 1988)]
than at distances further out sampled by the head of an expanding
radio-source later in its lifetime.  Where the alignment effect
is caused by dust-scattered quasar light, the certain
youthfulness of distant radio-galaxies alleviates the near
discrepancy (De Young 1998) between radio-source ages and the
time-scale for which dust grains can survive in the presence of
shocks caused by the advancing radio-jets.  The youth--redshift
degeneracy is consistent with the finding that the smallest
sources in a sample of $z \sim 1$ radio galaxies (all with very
similar luminosities) are those which are most aligned with
optical emission (Best et al 1996).  Best et al.\ remarked that
the sequence of changing optical aligned structure with
increasing radio size could be naturally interpreted by comparing
it with different phases of the interaction of the radio jets
with the interstellar and intergalactic media as the
radio-sources age.

{\bf Faraday depolarisation and the YZ degeneracy} Garrington \&
Conway (1991) have found a tendency for depolarisation to be
higher in sources at higher $z$.  Objects with higher $z$ which
are younger will be in much more recently merged environments
with the consequence that inhomogeneities in density or magnetic
field will more readily depolarise the synchrotron radiation from
the lobes.  Moreover, higher-$z$ sources being younger and
tending to be somewhat shorter will be closer in to the centre of
the potential well.  The higher density in this region will
enhance the observed depolarisation.

{\bf The YZ degeneracy and dust emission} Many of the highest-$z$
radio-galaxies have gas masses comparable to gas-rich spiral
galaxies (Dunlop et al.\ 1994; Hughes et al.\ 1998) and inferred
star-formation rates which, in the local Universe, are rivalled
only by galaxy-galaxy mergers like Arp 220 (Genzel et al 1998).
If high-$z$ objects are being viewed during a similar merging of
sub-components the associated star formation could be responsible
for a significant fraction of the stellar mass in the remnant
galaxy.  Since the high-$z$ radio-galaxies --- those which Hughes
et al.\ detected with SCUBA --- are necessarily young ($< 10^7$
years, see Fig.\ \ref{fig:tracks}), and since the whole merger
must take a few dynamical crossing times, or $10^{8-9}$ years,
the implication is that the event which triggered the
jet-producing central engine is synchronised with massive star
formation in a gas-rich system, perhaps as material streams
towards the minimum of the gravitational potential well of the
merging system.  The YZ degeneracy may help explain why few
lower-$z$ radio-galaxies show similarly large (rest-frame)
far-infrared luminosities compared to the high-$z$ population:
they are being observed significantly longer after the
jet-triggering event.

\section{Where are the dying classical doubles?}

Dead or relic classical doubles appear to be very rare.
(Giovannini et al.\ 1988, Harris et al.\ 1995, Harris et al.\
1993, Cordey 1987).  That there are so few implies that the
turn-off timescale, both for the cessation of jet-activity and
for the depletion of highly relativistic particles, must be very
rapid.

Once there is a cessation of jet-power, this implies that the
radio source luminosity will decrease into oblivion.  What will
govern this decline in luminosity?  The radio source will expand
causing adiabatic losses in the ways described in \S5.3 (see also
Scheuer \& Williams 1968), but the expansion of the plasma lobe
can only occur on timescales consistent with the sound speed so
this cannot be a rapid process.  A rapid decay time does however
come from considering the brevity of the radiative lifetimes of
the synchrotron emitting particles.

To radiate at rest-frame 5\,GHz in a magnetic field strength of
1\,nT, equation~\ref{eq:lorentz} tells us that the required
Lorentz factor of the electron is given by $\gamma \sim 10^4$.
The radiative lifetimes of synchrotron emitting particles are
often purported to be $\sim 10^{7-8}$ years, but such values
overlook the fact that the lobe of a radio source is an evolving
entity and the adiabatic expansion losses (especially the
decreasing magnetic field) discussed earlier mean that particles
with increasingly higher $\gamma$ are required to produce the
emission at the same chosen frequency as the source ages.  For
example, in the situation above, a $\gamma \sim 10^4$ particle
will emit most of its radiation at 5\,GHz if the magnetic field
strength is 1\,nT.  If the field lowers adiabatically to 0.1\,nT
then a particle whose Lorentz factor is $3 \times 10^4$ is
required to produce the emission at the same `observing
frequency' of 5 GHz.  In a radio lobe, changes in the magnetic
field go hand-in-hand with changes in the energies of the
particles themselves since it is the expansion of the lobe which
governs both.  Thus the particle now radiating in the lower
$B$-field of 0.1\,nT at 5\,GHz with $\gamma \sim 3 \times 10^4$
would, prior to the decrease in magnetic field strength, have had
$\gamma \sim 10^5$.  The synchrotron cooling rate increases as
$\gamma^2$, shortening the relevant radiative lifetimes further.

We explore quantitatively the radiative lifetimes of synchrotron
emitting particles in Blundell \& Rawlings (1999b), and find that
they appear to be substantially shorter than source ages (see
figure 1 of that paper).  If we accept that the synchrotron
radiative lifetimes are short then there is little difficulty
explaining why synchrotron radiation in the GHz regime should
cease on the same timescales.  There would thus be no problem
explaining the rarity of dead classical doubles.

But this picture leaves us with a problem in explaining how many
large linear-size, low-powered radio sources have lobes with
synchrotron emission often all the way back to the core.  These
sources are almost certainly the oldest examples of classical
doubles which we currently know of and at distances of many 100s
kpc from the hotspot, synchrotron emission is seen from particles
with a radiative lifetime substantially shorter than the age of
the source.  Reconciliation of these has traditionally come from
assuming a fast bulk backflow speed within the lobe which
transports the relativistic particles away from the hotspot
rather faster than the lobe itself advances away from the core.
But this brings us to another problem with this standard picture:
there must be a sink for all of this backflowing emission.  As
Jenkins \& Scheuer (1976) first pointed out, if synchrotron
losses play a major part in limiting lobe lengths, then lobes
lengths (and by extension backflow reservoirs) should be longer
and more extensive at low frequencies than at high frequencies.

However, we do not find this to be the case.  Images of bright 3C
sources made using the new 74\,MHz receivers on the VLA have
remarkably similar silhouettes to fully sampled images at
1.4\,GHz.  Some examples are shown in Fig.\ 11 illustrating this,
with images at GHz frequencies from the 3CRR Atlas Web-site
(Leahy, Bridle \& Strom 1996, {\tt
http://www.jb.man.ac.uk/atlas}) shown for comparison.  These
images show how the fractional contribution of the hotspots
changes as a function of frequency and especially that there is
no hint of any backflow emanating from the oldest parts of the
lobes: in all of the 3C sources we have observed to date there is
no evidence for any extended structure which is not already seen
at GHz frequencies.  A full set of these images together with
associated spectral index maps and analysis will be presented in
Kassim, Perley \& Blundell (in prep.).

\begin{figure}
\begin{center}
\includegraphics*[width=13.5cm]{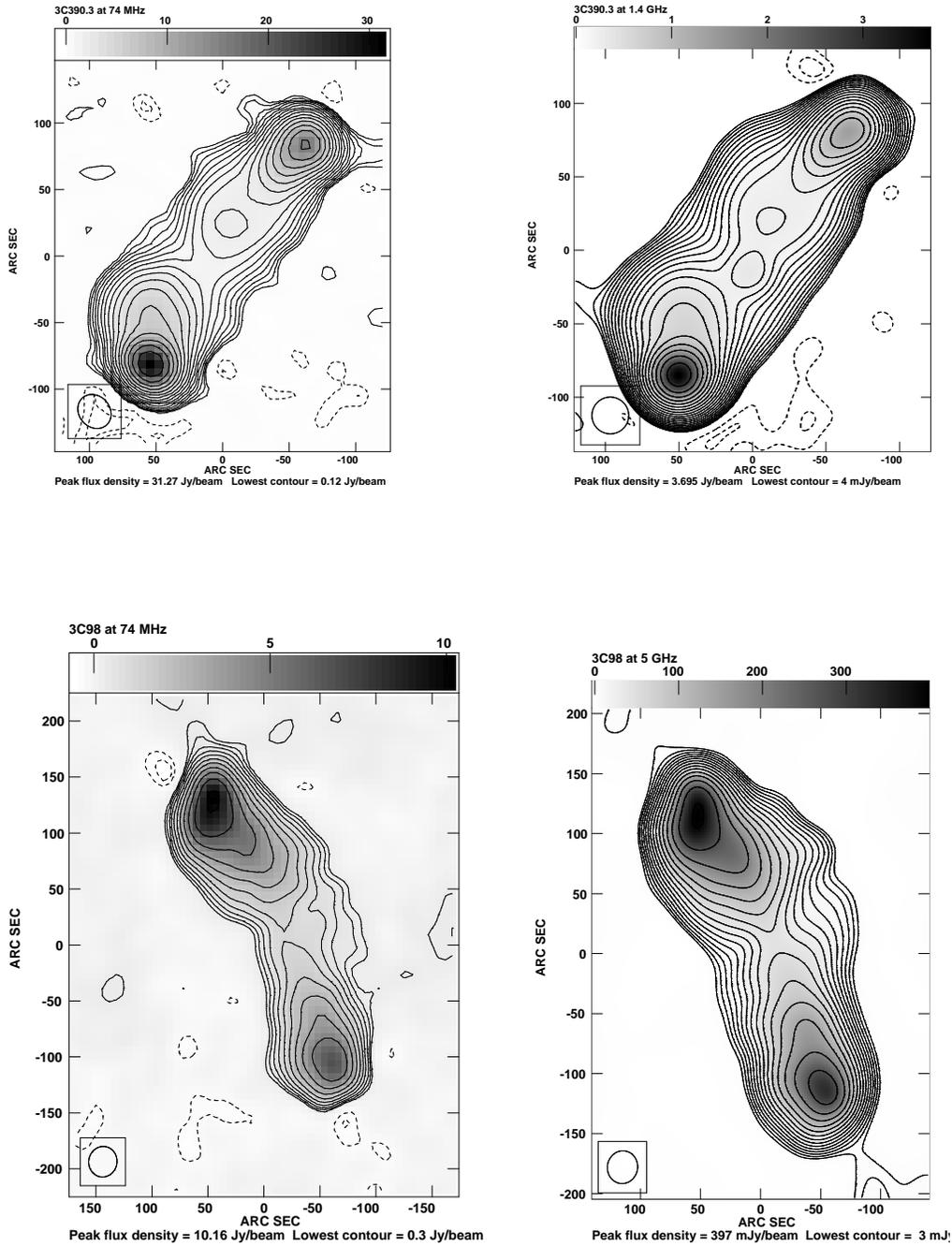}
\end{center}
\caption{Images at 74\,MHz and at GHz frequencies of 3C390.3 and
3C98.  Contours are logarithmic in intervals of $\sqrt{2}$.   }
\label{fig:4band}
\end{figure}

A rapid diffusion of the high Lorentz factor particles would
reconcile the rapid disappearance of classical doubles (note that
converting classical doubles into FRIs would be too slow a
process to solve this problem) and the presence of synchrotron
emitting material in lobes so far separated from the hotspot that
their radiative lifetimes are shorter than the
hotspot-travel-time from their location.  Particle
re-acceleration and inhomogeneous magnetic field strengths would
only solve the latter, but not the former, problem.  While it is
clear that classical diffusion is too slow a process, other
diffusion mechanisms which depend on the tangling scale and
strength of magnetic fields within the lobes (such as that
discussed by Duffy et al.\ 1995) merit further investigation.

\begin{flushleft}
{\bf Acknowledgements}
\end{flushleft}

K.M.B.\ thanks the Royal Commission for the Exhibition of 1851
for a Research Fellowship.  Basic research in radio astronomy at
the US Naval Research Laboratory is supported by the US Office of
Naval Research.  The VLA is a facility of the National Radio
Astronomy Observatory operated by Associated Universities, Inc.,
under co-operative agreement with the National Science
Foundation.

{\bf Peter Barthel:} Another explanation for the
$P$-$\alpha$ correlation would be enhanced radiative efficiency
in a hot confining medium (X-ray halo) such as is present around
Cygnus A and proposed to boost its radio luminosity (Barthel \&
Arnaud, 1996, MNRAS, 283, {\sc l}45).  Cosmologically evolving
halos (Gopal-Krishna \& Wiita, 1991, ApJ) plus the bias would
explain all the cosmological effects you attribute to youth.
But, I do like your youth model --- guess several mechanisms are
at work\ldots

{\bf Katherine Blundell:} It is true that a more dense confining
medium results in a higher radio luminosity (see Fig.\
\ref{fig:tracks}, which shows how an order of magnitude increase
in density will increase the radio luminosity by half an order of
magnitude).  It is not true that denser media will cause higher
spectral indices to be observed, and especially not for
low-frequency spectral indices.  It is worth reminding ourselves
that we are trying to explain an observed correlation of
luminosity with {\em low-frequency} spectral-index.  In fact the
reverse effect will be important: for a more dense medium the
expansion of the source is slower so the magnetic field does not
fall off so quickly.  In a {\em higher} magnetic field, lower
Lorentz factor particles are responsible for the radiation at a
fixed observing frequency.  Given the usual shape of energy
distributions in these objects (i.e.\ steep curved spectra),
lower Lorentz factor particles will give rise to {\em flatter}
not {\em steeper} spectra.  Thus this mechanism cannot be
responsible for the $P$--$\alpha$ correlation.

{\bf Paul Wiita:} How large was the sample of sources from which
   you concluded there was no evidence for backflow at 74\,MHz
   that wasn't seen at 1.4\,GHz?

{\bf Katherine Blundell:}  Just a dozen sources to date.  

{\bf Paul Wiita:} Some of the points you made concerning the
   youth-redshift degeneracy were made in a different context by
   Gopal-Krishna, myself and Saripalli (MNRAS, 1989, 239, 173)
   when we modelled the paucity of giant radio galaxies.

{\bf Katherine Blundell:} Our recent study which led to the
`youth-redshift degeneracy' has illustrated that in fact the
inverse Compton losses are not by any means the most important
factor in causing the deficit of large linear-size radio
galaxies.  It is actually the drastic effects of a survey {\em
flux-limit} together with luminosity tracks which decline because
of adiabatic expansion losses which predominantly reduces the
number of giant radio galaxies found in complete samples above $z
\sim 0.3$.  (It should be noted that giant radio galaxies out to
$z \sim 0.9$ have been found by Cotter, Rawlings \& Saunders
1996.)  Inclusion of the role of the hotspot also enables us to
explain not just the low-numbers of {\sl low}-powered large radio
galaxies, but also the problem posed by Baldwin (1982) that there
is an {\em absence} of {\sl high}-power large size sources.

\end{document}